\documentclass[RNAAS]{aastex62}
\usepackage{amssymb, amsmath,framed}

\usepackage{bm}
\expandafter\ifx\csname package@font\endcsname\relax\else
 \expandafter\expandafter
 \expandafter\usepackage
 \expandafter\expandafter
 \expandafter{\csname package@font\endcsname}
\fi
\def\bq{\begin{equation}}
\def\eq{\end{equation}}
\def\bqy{\begin{eqnarray}}
\def\eqy{\end{eqnarray}}

\begin{document}
\title{\large{Limitations of Chemical Propulsion for Interstellar Escape from Habitable Zones around Low-Mass Stars}}

\correspondingauthor{Manasvi Lingam}
\email{manasvi.lingam@cfa.harvard.edu}

\author{Manasvi Lingam}
\affiliation{Institute for Theory and Computation, Harvard University, Cambridge MA 02138, USA}
\affiliation{Harvard-Smithsonian Center for Astrophysics, Cambridge, MA 02138, USA}

\author{Abraham Loeb}
\affiliation{Institute for Theory and Computation, Harvard University, Cambridge MA 02138, USA}
\affiliation{Harvard-Smithsonian Center for Astrophysics, Cambridge, MA 02138, USA}

\section{}
On Earth, current and past space missions have been reliant on chemical rockets \citep{Craw90}. A notable issue with chemical rockets is that the mass of the fuel required (for a given payload) scales exponentially with the final speed that one wishes to achieve. Fortunately, this issue is not particularly important for humanity because the escape velocities from the Earth and the Solar system are roughly within an order of magnitude of the velocities achievable by chemical rockets. However, the use of chemical rockets for undertaking interstellar travel is likely to become more problematic for habitable planets around low-mass stars, for reasons outlined below. In this work, we carry out calculations to quantify, and further develop, the general points raised in \citet{Loeb18}. 

The rocket equation developed by \citet{Tsi03} states that
\begin{equation}\label{RockEq}
    \frac{m_0}{m_f} = \exp\left(\frac{\Delta v}{v_\mathrm{ex}}\right),
\end{equation}
where $m_0$ and $m_f$ are the initial (payload and fuel) and final (only payload) masses respectively, $v_\mathrm{ex}$ is the exhaust velocity and $\Delta v$ is the maximum velocity change that can be achieved. For state-of-the-art rockets that use liquid oxygen-hydrogen fuel, $v_\mathrm{ex} = g_\oplus I_\mathrm{sp}$ where $g_\oplus = 9.8$ m/s$^2$ is the Earth's surface gravity and $I_\mathrm{sp} \sim 450$ s is the specific impulse. If we wish to exit the planetary system, $\Delta v \gtrsim v_\mathrm{esc}$ is necessary, with the escape speed $v_\mathrm{esc}$ defined as
\begin{equation}\label{vesc}
    v_\mathrm{esc} = \sqrt{\frac{2 G M_\star}{a}},
\end{equation}
for a given stellar mass ($M_\star$) and semi-major axis of the planet ($a$). We have not considered the conditions for escaping the planet's gravity, which has been investigated in \citet{Hip18}. It can, however, be verified that the corresponding velocity for escaping the planetary surface is likely to be lower than $v_\mathrm{esc}$ for $M_\star \lesssim M_\odot$ if we consider rocky planets in the habitable zones of their host stars. In our discussion below, we will adopt the criterion $\Delta v > (1-1/\sqrt{2}) v_\mathrm{esc}$ instead, where the additional factor has been introduced to account for the boost from gravitational assists. This reduction in $\Delta v$ is achievable by launching the rocket such that it will be parallel to the planet's motion \citep{Loeb18}. In this scenario, the planet's speed in a circular orbit is $v_\mathrm{c} = \sqrt{G M_\star/a}$ and therefore $\Delta v = v_\mathrm{esc} - v_\mathrm{c}$ is given by $\Delta v = (1-1/\sqrt{2}) v_\mathrm{esc} \approx 0.29 v_\mathrm{esc}$. 

\begin{figure}
\includegraphics[width=7.5cm]{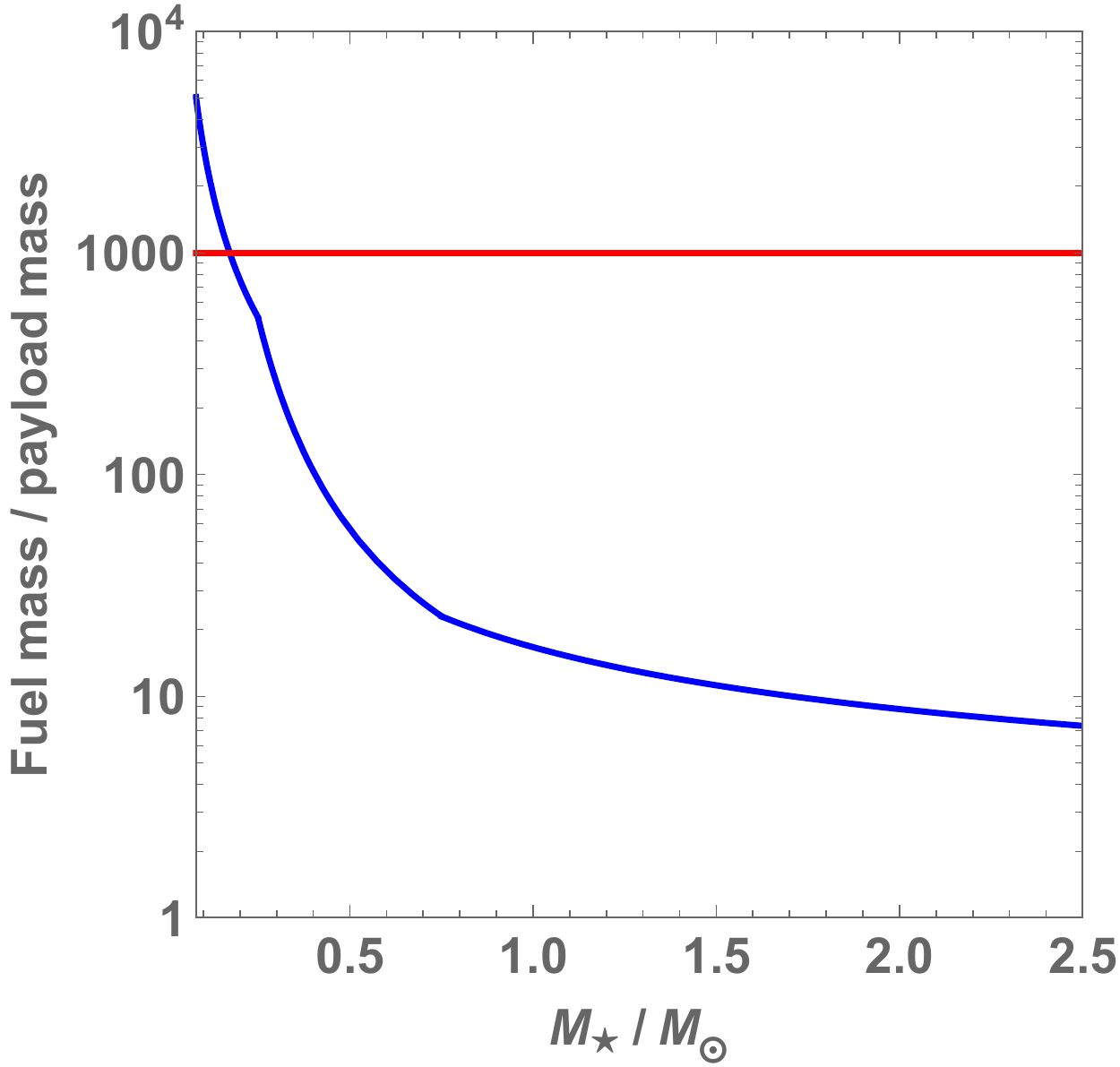} \\
\caption{The ratio of the initial mass (mostly fuel) to the final mass (payload) as a function of the stellar mass $M_\star$ based on (\ref{RockFin}). The red line represents the mass ratio cutoff at $\approx 10^3$, which corresponds to $M_\star \approx 0.2 M_\odot$.}
\label{Fig1}
\end{figure}

Although $v_\mathrm{esc}$ depends on two independent parameters, we can reduce this to just one parameter by demanding that the technological species inhabit an ``Earth-analog'', i.e. a habitable planet that receives the same insolation as the Earth so as to possess surface liquid water, which is necessary for the chemistry of ``life as we know it''.  In this event, we have $a \propto L_\star^{1/2}$ with $L_\star$ denoting the stellar luminosity. Thus, using the above data in (\ref{RockEq}), we find that the lower bound on $m_0/m_f$ is given by
\begin{equation}\label{RockFin}
    \frac{m_0}{m_f} = \exp\left[2.8 \times \left(\frac{M_\star}{M_\odot}\right)^{1/2} \left(\frac{L_\star}{L_\odot}\right)^{-1/4} \right],
\end{equation}
and we can make use of the mass-luminosity relation from \citet{LBS16} to express the right-hand-side purely as a function of $M_\star$. The final result has been plotted in Fig. \ref{Fig1}, and it is evident that $m_0/m_f$ becomes very high for low-mass M-dwarfs. We suggest that a ``reasonable'' cutoff for interstellar travel is given by $m_0/m_f \approx 10^3$. In this case, a payload with mass comparable to that of the Apollo mission ($\sim 45$ tons) would require a fuel mass that is one-fifth the mass of the Sears tower. This constraint imposes the limit $M_\star > 0.2 M_\odot$. 

In other words, it will not be easy for technological species orbiting low-mass M-dwarfs, such as Proxima Centauri and TRAPPIST-1, to undertake interstellar travel by means of chemical rockets. Consequently, these species may opt for alternative means of propulsion such as light sails, which have the added advantages of dispensing with carrying the fuel and achieving relativistic speeds.\footnote{Other notable propulsion methods capable of achieving high speeds include nuclear fusion engines and antimatter fuel \citep{Gil04}, but their utilization has not been demonstrated as of yet. Alternatively, ion (or tether) propulsion, perhaps coupled to the deployment of chemical rockets, may enable technological species around low-mass stars to undertake interstellar travel \citep{Ho18}.} Signatures of putative light sails could be detectable through either: (i) megastructures like Stapledon-Dyson spheres \citep{Dys60} for collecting stellar radiation, (ii) artificial spectral edges produced by photovoltaic arrays set up for tapping stellar energy \citep{LiLo17} or (iii) leakage of radiation from light sails \citep{GL15,LL17}.


\end{document}